\documentstyle[12pt]{article}
\def\beq{\begin{equation}}
\def\eeq{\end{equation}}
\def\bea{\begin{eqnarray}}
\def\eea{\end{eqnarray}}
\def\bq{\begin{quote}}
\def\eq{\end{quote}}
 
\begin{document}
\pagestyle{empty}
\begin{flushright}
{ROME prep.1223/98 \\
 hep-th/9810012}
\end{flushright}
\vspace*{5mm}
\begin{center}
{\bf SOME COMMENTS ON $S$-DUALITY IN FOUR-DIMENSIONAL $QCD$ }
\\  
\vspace*{1cm} 
{\bf M. Bochicchio} \\
\vspace*{0.5cm}
INFN Sezione di Roma \\
Dipartimento di Fisica, Universita' di Roma `La Sapienza' \\
Piazzale Aldo Moro 2 , 00185 Roma  \\ 
\vspace*{2cm}  
{\bf ABSTRACT  } \\

\end{center}
\vspace*{5mm}
\noindent
We show that a necessary condition, for the partition function of
four-dimensional Yang-Mills theory to satisfy a $S$-duality property,
is that certain functional determinants, generated by the dual change of 
variables, cancel each other. This result holds up to non-topological
boundary terms in the dual action and modulo the problem of field-strength
copies for the Bianchi identity constraint.
\vspace*{1cm}
\begin{flushleft}
September 1998
\end{flushleft}
\phantom{ }
\vfill
\eject

\setcounter{page}{1}
\pagestyle{plain}

\section{Introduction}

The recent upsurge of results about duality transformations in the
realm of string theories and supersymmetric gauge theories has raised
renewed interest in the role that duality transformations \cite{Ha1} might 
have in understanding or even solving non-supersymmetric gauge theories
such as $QCD$.
This paper is essentially just a comment, with perhaps a contemporary bias,
to an `old' paper of Mecklenburg and Mizrachi \cite{MM} that studied
duality transformations in Yang-Mills theories, in an attempt to extend to the
non-abelian Yang-Mills case \cite{H} exact $S$-duality results for abelian
systems of charges and monopoles in two and four dimensions \cite{C,Ca}.
Ref. \cite{MM} is itself the extension to the case of non-zero $\theta$
angle of Halpern \cite{Ha1} pioneering work on non-abelian duality
in four-dimensional $YM$ theories.
Mecklenburg and Mizrachi established a formula for the action of
a duality transformation in $YM_4$, that showed that the dual theory, in 
addition
to exchanging weak with strong coupling, possesses new topological and
non-topological boundary terms, that where instead absent in the
original action. They argued, in addition, that the non-topological terms
become irrelevant at weak coupling and in the semiclassical
approximation.\\
The main point of this paper, that is perhaps just a comment on the paper of
Ref. \cite{MM}, is that, in addition to the surface terms taken in consideration
in Ref. \cite{MM}, the duality transformation generates also some functional 
determinants that are implicit in Ref. \cite{MM}, but that do not seem
to have been taken explicitly into consideration.
The question then arises, as to whether they cancel each other, in such a 
way that the dual action has the same functional dependence from the dual
fields as the original one, up to the mentioned boundary terms.
An affirmative answer would imply, in modern terms, $S$-duality for the
four-dimensional $YM$ partition function (up to the non-topological boundary 
terms, but a certain degree of freedom in integrating by parts seems common
to all the duality transformations in the literature).
Yet, we should mention that the present discussion ignores the problem
of field-strength copies \cite{Ha2},\cite{SS}, that may affect the solution of
the Bianchi identity constraint in terms of the dual field.
The details are described in the following section.

\section{The duality transformation}

We describe the duality transformation according to Ref. \cite{MM}.
The Euclidean partition function of $YM_4$ with a $\theta$-term is
defined by the formula:
\bea
Z&=&\int \exp[- \frac{1}{4g^2} \int (F^2 -i \tilde{\theta} F \tilde{F})] DA 
\nonumber \\ 
\tilde{\theta}&=& \frac{g^2}{8\pi^2} \theta  \nonumber \\
F_{\mu \nu}&=&
\partial_{\mu}A_{\nu}-\partial_{\nu}A_{\mu}+i [A_{\mu},A_{\nu}] \;,
\eea             
where the dual field strength is:
\bea
\tilde{F}_{\mu \nu}&=&
 \frac{1}{2}\epsilon_{\mu \nu \alpha \beta}  F_{\alpha \beta}
\eea 
and the trace and sum over the Euclidean indices are understood.
Introducing the auxiliary variable $K$, the partition function can be 
written as a Gaussian integral over $K$:
\bea
Z&=&\int \exp[-\int \frac{g^2}{4(1+\tilde\theta^2)}( K^2+i \tilde{\theta} K
 \tilde K)+\frac{i}{2}\tilde{K} F(A)] DA DK \; .
\eea
After integrating by parts, the $A$ integral becomes Gaussian.
The dual field, $A^D$, is introduced in the following way.
The quadratic part in $A$ of the functional integral is represented
as another Gaussian integral, involving the auxiliary field $B$.
After the shift $B'=B+\xi_{\Sigma}$, the field redefinition
$B'=ad_{\tilde{K}} \cdot A^D$ and some integration by parts, the partition 
function becomes:
\bea
Z &=& \int \exp[-\int \frac{g^2}{4(1+\tilde{\theta}^2)}( K^2+i 
\tilde{\theta} K \tilde{K})+\frac{i}{2} \tilde{K} F(A^D)
 -\frac{i}{2} \xi_{\Sigma}(ad_{\tilde{K}})^{-1} \xi_{\Sigma}] \times
 \nonumber \\
&& \times
|Det(ad_{\tilde{K}})|^{\frac{1}{2}}\delta(d^*_{A^D} \tilde{K}) DA^D DK \; ,
\eea
where $ad_{\tilde{K}} \cdot$ denotes the adjoint action of $\tilde{K}$ 
and $K$ carries, as a multiplication operator, one-forms into one-forms. The 
delta-function
is obtained integrating over the $A$ field.
The $\xi_{\Sigma}$ term is a non-topological boundary term given by:
\bea
\int \xi_{\Sigma} V &=& \int \partial_{\mu} (\tilde{K}_{\mu \nu} V_{\nu})
\eea
for any $V$.
The delta-functional constraint is the Bianchi identity, that is solved by:
\bea
K&=&F(A^D)  \; .
\eea
This solution of the constraint is not the most general solution 
\cite{Ha2}, \cite{SS}.
We ignore this difficulty for the moment, assuming that the solution in terms 
of the field strength of the dual field is generic.
After solving for the Bianchi identity, the dual action has the same
functional dependence from the dual fields as the original one,
but with new coupling constants, that exchange the weak with the strong 
coupling, and with new boundary terms. Some of them have a topological
meaning and are equivalent to an extra shift in the $\theta$ angle.
We now come to the point of our paper. The field redefinition
has generated an extra functional determinant, that it was not
present in the original functional integral.
However, this functional determinant must be combined with another
determinant, that comes from solving the delta-functional 
constraint associated to the Bianchi identity. It is formally given by:
\bea
|Det(d^*_{A^D})|^{-1} \; ,
\eea
where the operator in the determinant carries antisymmetric two-forms
into one-forms and the associated determinant must be suitably interpreted
\cite{S}.
In this way we get, for the final form of the partition function as a 
functional integral over the dual variables, $A^D$:
\bea
Z &=& \int \exp[-\int \frac{g^2}{4(1+\tilde{\theta}^2)}(F(A^D)^2+i 
\tilde{\theta}  F(A^D) \tilde{F}(A^D))+ \nonumber \\
&& + \frac{i}{2} \tilde{F}(A^D) F(A^D) 
-\frac{i}{2} \xi_{\Sigma} (ad_{\tilde{F}(A^D)})^{-1} 
\xi_{\Sigma}] \times \nonumber \\
&& \times
|Det(ad_{\tilde{F}(A^D)})|^{\frac{1}{2}} |Det(d^*_{A^D})|^{-1} DA^D \; .
\eea
If the two determinants cancel each other, the partition function of $QCD$
satisfies a $S$-duality identity, up to the boundary terms in the dual action
and modulo the field-strength copies problem:
\bea
Z(g,\theta)&=& const \times Z(g^D,\theta^D) \; ,
\eea
where the constant may depend from the couplings. The dual coupling
constant and the dual $\theta$-angle are easily read from Eq. (8).
If they do not cancel, the dual action is not the integral of a
local density and no (simple) $S$-duality holds for the $QCD$ partition
function.


\begin{thebibliography}{99}
\bibitem{Ha1} M. B. Halpern, {\it Phys. Rev.} {\bf D 16} (1977) 1798. \\
              M. B. Halpern, {\it Phys. Rev.} {\bf D 19} (1979) 517.
\bibitem{MM} W. Mecklenburg, L. Mizrachi, {\it Phys. Rev. D} {\bf D 27,8}(
1983) 1922.
\bibitem{H} G. 't Hooft,  {\it Nucl. Phys.} {\bf B 190 (FS 3)} (1981) 455.
\bibitem{C} J. L. Cardi and E. Rabinovici,  {\it Nucl. Phys.} {\bf B 205 (FS 5)}
 (1982) 1.
\bibitem{Ca} J. L. Cardi,  {\it Nucl. Phys.} {\bf B 205 (FS 5)}
 (1982) 17.
\bibitem{Ha2} M. B. Halpern, {\it Nucl. Phys.} {\bf B 139} (1978) 477.
\bibitem{SS} P. Majumdar and H.S. Sharatchandra, {\it` Duality 
transformations for $3+1$ dimensional Yang-Mills theory '} hep-th/9805102 
and references therein.
\bibitem{S} A.S. Schwarz,  {\it Comm. Math. Phys.} {\bf 67}
 (1979) 1.
\end{thebibliography}
\end{document}